\documentclass{jetpl}
\twocolumn

\usepackage[dvipsnames]{xcolor}
\usepackage{amsmath}
\usepackage{bm}
\usepackage{graphicx}
\usepackage{xcolor}
\usepackage{hyperref}
\usepackage{float}
\usepackage[numbers,square,comma]{natbib}
\newcommand{\change}[1]{\textcolor{black}{#1}}

\newcommand{\newchange}[1]{\textcolor{black}{#1}}

\lat

\title{Universal bottleneck for thermal relaxation in disordered metallic films}

\rtitle{Universal bottleneck for thermal relaxation in disordered metallic films}

\sodtitle{Universal bottleneck for thermal relaxation in disordered metallic films}

\author{E.\,M.\,Baeva$^{\times,+}$, N.\,A.\,Titova$^{+}$, A.\,I.\,Kardakova$^{+,*}$, 
	S.\,U.\,Piatrusha$^\times$, V.\,S.\,Khrapai$^{\times,+}$\/\thanks{e-mail: dick@issp.ac.ru}}

\rauthor{E.\,M.\,Baeva, N.\,A.\,Titova, A.\,I.\,Kardakova, S.\,U.\,Piatrusha, V.\,S.\,Khrapai}

\sodauthor{Baeva, Titova, Kardakova,  Piatrusha, Khrapai}

\address{$^\times$Institute of Solid State Physics, Russian Academy of Sciences, 142432 Chernogolovka, Russia}
\address{$^+$Moscow State University of Education, 29 Malaya Pirogovskaya Street, Moscow 119435, Russia}
\address{$^*$National Research University Higher School of Economics, 20 Myasnitskaya Street, Moscow 101000, Russia}

\dates{\today}{*}

\abstract{
	We study the heat relaxation in current biased metallic films in the regime of strong electron-phonon coupling. A thermal gradient \change {in the direction normal to} the film is predicted, \change{with a spatial temperature profile determined by the temperature-dependent heat conduction. In the case of strong phonon scattering the heat conduction occurs predominantly via the electronic system and the profile is parabolic.} This regime leads to the linear \change{dependence of the} noise temperature as a function of voltage bias, \change{in spite of the fact that all the dimensions of the film are large compared to the electron-phonon relaxation length. This is in stark contrast to the conventional scenario of relaxation limited by the electron-phonon scattering rate. A} preliminary experimental study of \change{a 200\,nm thick} NbN film \change{indicates} the relevance of our model \change{for materials used in } superconducting nanowire single-photon detectors.
}


\begin{document}

\maketitle

Kinetics of the energy relaxation defines a temporal response of radiation detectors based on thin metallic films at low temperatures~\citep{Natarajan2012,holzman2019,Marsili2016,Zhang2018,Klapwijk2017,Tamireaau3826}. According to advanced theoretical models~\citep{Klapwijk2017,Vodolazov2017}, the internal efficiency of the superconducting nanowire single-photon detectors (SNSPDs) depends critically on the time-scales of the electron-phonon (e-ph) relaxation $\tau_{\rm e-ph}$ and of the phonon escape in the substrate $\tau_{\rm esc}$. Qualitatively, the $\tau_{\rm e-ph}$, along with the diffusion coefficient of electrons, determines the characteristic size of the hot-spot mediated by the absorption of a photon, whereas the $\tau_{\rm esc}$, if big enough, can limit the eventual relaxation time of the hot-spot~\citep{Annunziata2010,Marsili2011,Zhang2019}. In addition, important is the ratio of the electron and phonon heat capacities, $C_{\rm e}/C_{\rm ph}$, which defines the fraction of the photon energy going into the electron system~\citep{Vodolazov2017,Baeva2018}. 

A correspondence of several microscopic time-scales and the relaxation time, usually probed in experiments on the amplitude modulated absorption of the radiation~\citep{Rall2010,kardakova2013,Sidorova2018,sidorova2019electron}, is not straightforward. The interpretation of such experiments invokes the models of the energy balance of varying complexity. In some cases, the inclusion of a specific relaxation bottleneck in the model~\citep{Sidorova2018,sidorova2019electron} can explain the relaxation times much longer than $\tau_{\rm e-ph}$, in spite of a similar temperature dependence. The problem is even more intricate since the materials suitable for SNSPDs typically have $C_{\rm e}/C_{\rm ph}\gtrsim1$ near the superconducting transition temperature~\citep{Vodolazov2017}. This, by the detailed balance, is equivalent to $\tau_{\rm e-ph}\gtrsim\tau_{\rm ph-e}$, where $\tau_{\rm ph-e}$ is the phonon re-absorption time by the electrons. The latter inequality is, essentially, the condition of \change{the} strong coupling regime of the electron and phonon systems, which manifests in local thermal equilibration between them when $\tau_{\rm esc}\gg\tau_{\rm ph-e}$. In this regime,  one would not expect the microscopic time-scales $\tau_{\rm e-ph}$ and $\tau_{\rm ph-e}$ to govern the relaxation process individually.

In this article we focus on the heat transport in a current biased disordered metallic film in the regime of strong coupling between the electrons and phonons. We investigate the thick film limit, in which the mean-free paths of both the electrons and phonons are small compared to the film thickness. In this case, the heat outflow in the substrate requires a thermal gradient transverse to the film, with the spatial profile determined by the temperature dependence of the total thermal conductivity. Remarkably, in a situation when the electron contribution to the heat conductivity dominates, the spatial profile is parabolic and insensitive to the parameters of the electron-phonon relaxation. Here, we predict a non-vanishing shot noise of the film with a Fano factor $F=\sqrt{3}/2(d/l)$, which is determined solely by the ratio of film thickness $d$ and the length of the device $l$. Such a universal expression emphasizes the fact that the bottleneck for thermal relaxation in this case is imposed by the Wiedemann-Franz heat conduction transverse to the film, while the electron-phonon parameters drop out. Our preliminary measurements in a disordered $200$\,nm thick NbN film in the normal state are consistent with this result.

\begin{figure}[ht]
	\begin{center}
		\includegraphics[scale=1]{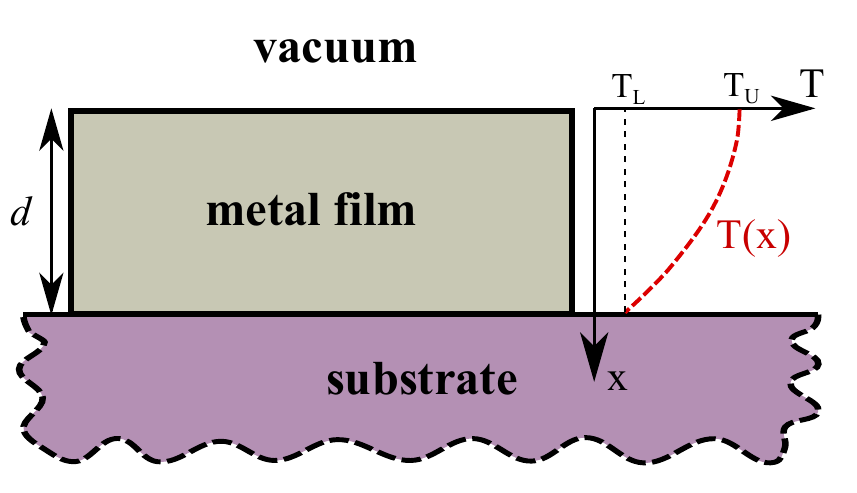}
		\caption{FIG. \ref{fig1}. The sketch of the experimental configuration under discussion. A conducting film with thickness $d$ is located on a substrate and heated uniformly via Joule heat. Heat relaxation through the heat conduction to the substrate results in a temperature gradient transverse to the film in the direction $x$. The axes demonstrate parabolic temperature profile $T(x)$ for the case of electron heat conduction according to Wiedemann-Franz law being dominant in the film. The ($T_{\rm U}$) and ($T_{\rm L}$) are the temperatures on the upper and lower surfaces, respectively.}\label{fig1}
	\end{center}
\end{figure}

We consider a typical experiment on the heat relaxation in a metallic film on a substrate. The electric current $I$ is biased  through the film with thickness $d$, length $l\gg d$, width $w\gg d$ and conductivity $\sigma$ (see Fig.~\ref{fig1}). The current distribution is uniform across the film section so that the current density $j=I/(wd)$. For the experimentally preferred case of a film much longer than the electron-phonon relaxation length $l_{\rm e-ph} \ll l$, we approximate the heat dissipation to occur entirely through the phonon conduction to the substrate. We also consider the case of strongly coupled electron and phonon subsystems, which allows us to define a single local equilibrium temperature $T$. This assumption requires the film thickness much larger than both electron-phonon and phonon-electron relaxation lengths $l_{\rm e-ph},\,l_{\rm ph-e} \ll d$. As such, we neglect the in-plain $T$ gradient and allow the temperature to be a function of a transverse coordinate $x$ (see axes on fig.~\ref{fig1}). We note that the temperature gradient between the upper and lower surfaces of the film is not addressed in most thermal relaxation studies, as it is either the electron-phonon coupling or the phonon thermal resistance at the interface that is considered to be the process limiting the heat transfer~\citep{Sidorova2018}.

The regular thermal balance equation:

\begin{equation}
	-\frac{\partial}{\partial x}\left(\kappa\frac{\partial}{\partial x} T\right) = \sigma^{-1}j^2,
	\label{heat_balance}
\end{equation}
sets the dissipation of Joule heat $\sigma^{-1}j^2$ via combined electron and phonon thermal conductivity $\kappa=\kappa_{\rm e}+\kappa_{\rm ph}$. Obviously, the eq.~(\ref{heat_balance}) implies a transverse temperature gradient with a spatial profile determined by the functional dependence $\kappa(T)$. Below we concentrate on a special case of negligible $\kappa_{\rm ph}$, which is plausible in extremely disordered metallic films. Here, similar to amorphous materials~\citep{Zeller1971}, we expect the $\kappa_{\rm ph}$ to cut-off at increasing temperature owing to the Rayleigh scattering, which gives rise to a fast decay of the mean free path of the acoustic phonon $l_{\rm ph}\propto \omega^{-4}$ as a function of its frequency $\omega$. The electron thermal conductivity itself is given by the the Wiedemann-Franz law $\kappa_{\rm e}=\mathcal{L}T\sigma$, where $\mathcal{L}=\pi^2k_B^2/3e^2$ is the Lorenz number. This leads to the standard parabolic solution for $T(x)$ ~\citep{Nagaev1995}:

\begin{equation}
	T^2(x)=T_{\rm L}^2+\left(T_{\rm U}^2-T_{\rm L}^2\right)\left(1-\frac{x^2}{d^2}\right), \label{parabola}
\end{equation}
where temperatures at the upper and lower surfaces of the film are denoted as $T_{\rm U}\equiv T(x=0)$ and  $T_{\rm L}\equiv T(x=d)$, respectively. Note that here we took into account the boundary condition of zero heat flux on the upper surface, which is assumed to be placed in vacuum, see Fig.~\ref{fig1}.
%
The substitution of solution (\ref{parabola}) into the equation (\ref{heat_balance}) leads to the relation:

\begin{equation}
	T_{\rm U}^2-T_{\rm L}^2=\frac{j^2d^2}{\mathcal{L}\sigma^2}.\label{solution}
\end{equation}
This solution satisfies the second boundary condition for the heat flux on the lower surface, namely that the heat flux density coincides with the Joule heat dissipated per unit area of the film.

As a next step, we presume, that the phonon escape into the bulk \change{of the substrate} provides an efficient path of heat relaxation in thin ($\sim l_{\rm ph-e}$) layer near the lower surface of the film, so that $T_{\rm L}\approx T_{\rm bath}$. \change{Hence}, considering the case of an intense heating $T_{\rm U} \gg T_{\rm bath}$ we get the solution $T_{\rm U}=jd/\sigma\mathcal{L}^{1/2}$.

The linear dependence $T_{\rm U}\propto I$ is reminiscent of the shot noise behavior in metallic diffusive conductors in the absence of electron-phonon relaxation~\citep{Nagaev1995,Kozub1995}. The same qualitative behaviour is the case for the noise temperature $T_{\rm N}$ of the whole sample. The $T_{\rm N}$ is determined by the average temperature, weighed by a local Joule power~\citep{Piatrusha2017}, which in present case corresponds to simple spatial averaging:

\begin{equation}
	T_{\rm N}=d^{-1}\int T(x)dx\approx \frac{\sqrt{3}}{4}\frac{ejd}{k_{\rm B}\sigma}=\frac{\sqrt{3}}{4}\frac{d}{l}\frac{eV}{k_{\rm B}},\label{Tnoise}
\end{equation}
where $V=lI/dw\sigma$ is the applied voltage bias. Comparing the result (\ref{Tnoise}) with the classical solution for a metallic diffusive conductor cooled via electron thermal conductivity through the contacts~\citep{Nagaev1995,Kozub1995}:
$T_{\rm N}=\sqrt{3}eV/2k_{\rm B}$, we observe that the strong electron-phonon coupling leads to the \change{drastic} reduction of a noise temperature by a geometrical factor $2l/d\gg1$. \change{Remarkably, while the suppression of the shot noise in the latter case  is expected~\citep{Nagaev1992}, the linear dependence $T_{\rm N}\propto V$ in the limit $l,d,w\gg l_{\rm e-ph}$
is special and manifests the relaxation bottleneck provided by the Wiedemann-Franz heat conduction transverse to the film.}
Finally, we give a rigorous expression for the $T_{\rm N}$, valid for arbitrary relation between the  $T_{\rm U}$ and $T_{\rm L}$ and obtained via integration of (\ref{parabola}):

\begin{equation}
	T_{\rm N}=\frac{1}{2}\left(T_{\rm L} + \frac{T_{\rm U}^2}{\sqrt{T_{\rm U}^2-T_{\rm L}^2}}\arccos\frac{T_{\rm L}}{T_{\rm U}}\right).
	\label{exact}
\end{equation}

In the following we concentrate on a preliminary experimental study of thermal relaxation in a $d=200$\,nm thick disordered NbN film. This material choice is natural for the purpose of reaching the regime of strong electron-phonon coupling. Much thinner NbN films of similar quality are routinely used in SNSPDs \change{~\citep{Baeva2018, Smirnov_2018}} and characterized by $\tau_{\rm e-ph}\sim\tau_{\rm ph-e}\sim10$\,ps near the superconducting transition temperature around $T\approx10$\,K. This very well meets the above criterion $l_{\rm e-ph},\,l_{\rm ph-e} \ll d$. The situation with phonon heat conductivity is more ambiguous~\citep{Sidorova2018}, yet we expect the Rayleigh scattering of the acoustic phonons to become a limiting factor in this highly disordered material, at least  at higher temperatures used in present experiment, such that $\kappa_{\rm ph}\ll\kappa_{\rm e}$.

The NbN film was grown the SiO$_2$-Si substrate at a room temperature using a DC magnetron sputtering system with $99.9999$\% pure niobium target. The preliminary vacuum was $7\cdot 10^{-7}$\,Torr. During the deposition, the magnetron power was fixed at $200$\,W and the gas mixture ratio was ${\rm Ar:N_2} = 40:7$. \change{In this way a strongly disordered NbN film was obtained with a room temperature resistivity of about 800\,$\rm \mu\Omega\cdot cm$ measured by a van der Pauw method. The} NbN film was further patterned \change{in a bridge of} a width of $w=0.99$\,$\mu$m and a length of $l=27.5$\,$\mu$m. \change{Prior to patterning, the} metal Ti/Au \change{(5\,nm/200\,nm)} contact pads were fabricated with a lift-off method \change{by means of} electron-beam lithography and thermal evaporation. \change{Next}, a $250$-nm thick protective Al mask was formed on the sample using the electron beam lithography and the electron beam evaporation. \change{Finally}, the film was etched in a mixture of  Ar and SF$_6$ gases \change{and } the aluminum mask removed in a KOH solution. 

The experimental setup for noise thermometry was built in a homemade liquid $^4$He insert, with a tank circuit at a resonance frequency of $40$\,MHz \change{at the input of a vapor-cooled high impedance low-noise amplifier ($\approx6$\,dB  gain  and $\approx3\times10^{-27}$\,A$^2$/Hz input current noise)}. The signal is further amplified by a chain of low-noise amplifiers at $300$\,K, filtered and measured via a power detector (see Supplemental Material of ~\citep{TikhonovPRB2014} for the details on shot-noise measurement technique \textcolor{black}{and Ref.~\citep{Piatrusha2018} for a recent review}).

In this experiment, the voltage bias is applied to the NbN sample causing the Joule heating of the electron subsystem and the subsequent current noise increase. The noise temperature $T_{\rm N}$ is obtained from the Johnson-Nyquist relation $S_{\rm I} = 4k_B T_{\rm N}/R$.

\begin{figure}[ht]
	\includegraphics[width = 0.95\linewidth]{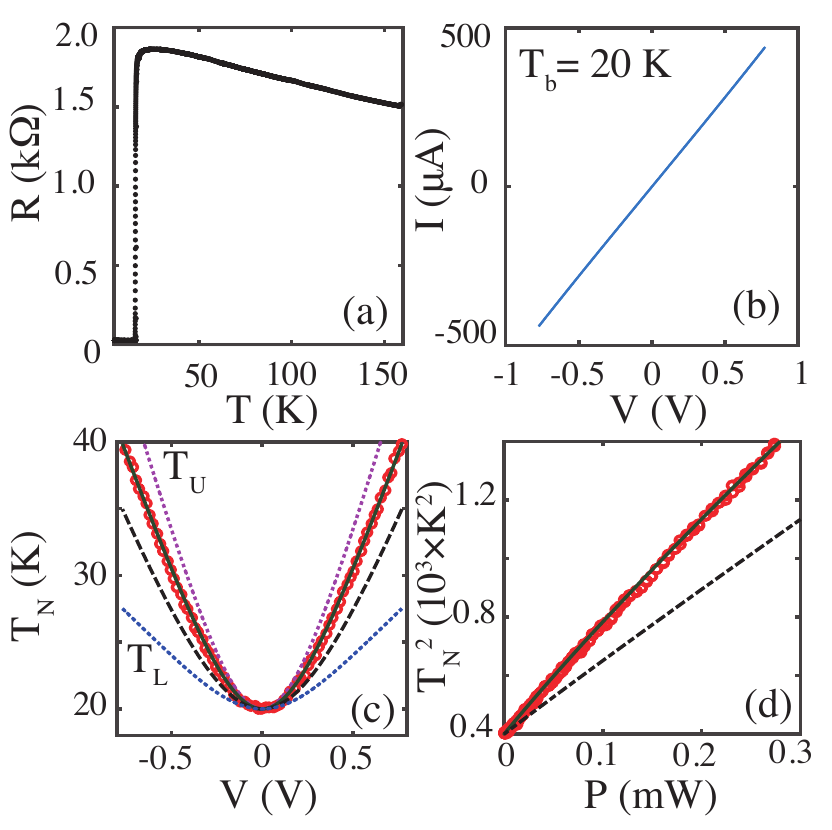}
	\caption{FIG. \ref{fig2}. The experimental data for 200-nm thick NbN sample. The sample goes to the superconducting state at $16$\,K, the width of the superconducting transition is about $2$\,K. (b) The I-V curve of the sample in the normal state at $T_b =20$\,K. (c) The noise temperature $T_N$ as a function of voltage bias $V$. The symbols are experimental data, the dashed line is the model result \newchange{under the assumption of perfect thermal coupling between the film and the substrate and $T_{\rm L}=T_{\rm bath}$. The difference between the experiment and the model can be improved provided an additional relaxation bottleneck is taken into account in the form of Kapitza resistance (see text). The corresponding dependencies of $T_{\rm N}$ and $T_{\rm L},\,T_{\rm U}$ are shown, respectively, by the solid line and two dotted lines.} (d) The same data for the noise temperature as in (c) plotted as $T^2_{\rm N}$ in dependence of the Joule power $P$. The data supports the relevance of the Wiedemann-Franz bottleneck model for thermal relaxation (see text).}\label{fig2}
\end{figure}

The panel Fig.~\ref{fig2}a demonstrates the linear response resistance of the sample \change{plotted} as a function of the bath temperature. A superconducting transition occurs at $T_{\rm C}= 16\,$K. On the panel Fig.~\ref{fig2}b the I-V curve is shown for $T_{\rm bath}=20\,$K well above the $T_{\rm C}$, which is linear up to $5\%$ in all measurement range. These data demonstrate a conventional low-temperature transport response of our NbN sample.

The main experimental result is presented on the panel Fig.~\ref{fig2}c. The symbols are experimental measurements, while the dashed line is the noise temperature \change{according to the equations (\ref{solution}) and (\ref{exact})} with the corresponding $l,\,d$ and $T_{\rm L}=T_{\rm bath}$. \change{As expected, the} noise temperature is low compared to the case of cooling via contacts (not shown), \change{evidencing a} strong thermal relaxation. At the same time, \change{both the absolute value of the $T_{\rm N}$ and its functional dependence on the bias voltage $V$ are close to our prediction}. Fig.~\ref{fig2}d \change{presents} the same data in \change{the} form of $T^2_{\rm N}$ as a function of \change{the} Joule power $P$, \change{which also follows from our model. Again, we observe} that the main heat relaxation bottleneck in this experiment is close to the one expected from the Wiedemann-Franz \change{heat conduction transverse to} the film. 

\newchange{The magnitude of the $T_{\rm N}$ growth with $V$ is stronger in the experiment as compared to the model. We attribute this difference to the effect of an additional process weakly limiting the thermal relaxation in the film or in the substrate. In the following we demonstrate how one such possible mechanism, namely the Kapitza resistance owing to the acoustic mismatch between the film and the substrate, could explain the experimental data. In this scenario, the temperature at the lower surface of the film exceeds the bath temperature $T_{\rm L}>T_{\rm bath}$ such that $P/A=A_{\rm K}(T_{\rm L}^4-T_{\rm bath}^4)$, where $A=wl$ is the area covered by the film and $A_{\rm K}\sim 100\div1000\,{\rm Wm^{-2}K^{-4}}$ is the parameter of the acoustic mismatch model~\citep{Elo2017}. Using the value of $120\,{\rm Wm^{-2}K^{-4}}$ we obtained the dependencies of the $T_{\rm L}$ and $T_{\rm U}$ on the bias voltage, which, via the eq.~(\ref{exact}), closely describe the experimental dependence of the noise temperature $T_{\rm N}$.  In Fig.~\ref{fig2}c these dependencies are shown, respectively, by the lower and upper dotted lines and by the solid line (the latter is also shown in Fig.~\ref{fig2}d). In spite of a substantial increase of the $T_{\rm L}$ above the bath temperature, caused by the Kapitza resistance, a strong temperature gradient transverse to the film is the case. Still, we would like to stress that in order to draw a definite conclusion an independent measurement of the substrate-mediated relaxation bottleneck is necessary, which goes beyond the scope of this work.}


In summary, we provide a model of thermal relaxation in disordered metallic films \change{in the regime of strong} electron-phonon coupling. We predict a sizeable temperature gradient transverse to the current-biased film, with a spatial profile determined by the heat conduction of the material. In the limit of dominant heat conduction via electrons the temperature profile is parabolic and the noise temperature of the film scales linearly with the bias voltage. This resembles a universal shot noise behavior in diffusive conductors with negligible electron-phonon coupling, yet with the noise temperature drastically suppressed by the geometrical factor $2l/d\gg1$. A preliminary  experiment in a thick strongly disordered NbN film is \newchange{not far from} our model predictions.


We acknowledge valuable discussions with I.V.~Tretyakov and A.V.~Semenov. The theoretical model was developed with a support from the RFBR project 19-32-80037. The fabrication of the NbN sample and transport characterization were supported by the RSF project 17-72-30036. 
Noise measurements were performed with a support from the RSF project 19-12-00326. A.I.K. and E.M.B acknowledge financial support under the Grant of the President RF MK-1308.2019.2. The data analysis was performed within the state task of the ISSP RAS. 

\bibliographystyle{unsrtnat}
\bibliography{refWF}

\end{document}